\documentclass[twocolumn,prl]{revtex4-1}
\usepackage[latin9]{inputenc}
\usepackage{color}
\usepackage{amsmath}
\usepackage{amssymb}
\usepackage{graphicx}
\usepackage{esint}
\usepackage[unicode=true,pdfusetitle,
 bookmarks=false,
 breaklinks=false,pdfborder={0 0 0},backref=false,colorlinks=true]
 {hyperref}
\hypersetup{
 colorlinks,linkcolor=red,citecolor=blue}
\usepackage{breakurl}

\makeatletter

\usepackage{amsfonts}

\makeatother

\begin{document}

\title{Multimode Strong Coupling in Superconducting Cavity Piezo-electromechanics}

\author{Xu Han, Chang-Ling Zou, and Hong X. Tang}

\email{hong.tang@yale.edu}

\address{Department of Electrical Engineering, Yale University, New Haven,
Connecticut 06511, USA}
\begin{abstract}
High frequency mechanical resonators subjected to low thermal phonon
occupancy are easier to be prepared to the ground state by direct
cryogenic cooling. Their extreme stiffness, however, poses a significant
challenge for external interrogations. Here we demonstrate a superconducting
cavity piezo-electromechanical system in which multiple modes of a
bulk acoustic resonator oscillating at $10\,\textrm{GHz}$ are coupled
to a planar microwave superconducting resonator with a cooperativity
exceeding $2\times10^{3}$, deep in the strong coupling regime. By
implementation of the non-contact coupling scheme to reduce mechanical
dissipation, the system exhibits excellent coherence characterized
by a frequency-quality factor product of $7.5\times10^{15}\,\textrm{Hz}$.
Interesting dynamics of temporal oscillations of the microwave energy
is observed, implying the coherent conversion between phonons and
photons. The demonstrated high frequency cavity piezo-electromechanics
is compatible with superconducting qubits, representing an important
step towards hybrid quantum systems.
\end{abstract}
\maketitle
\emph{Introduction.-} Controlling mechanical motion by electromagnetic
field has recently emerged as a new field of great interest \citep{Tsang2010,Regal2011,Andrews2014}.
Impressive progresses have been achieved in various cavity electro/optomechanical
systems, including ground-state cooling of mechanical resonators \citep{Teufel2011a,Chan2011,Peterson2016},
quantum squeezing of a mechanical mode \citep{Wollman2015,Lecocq2015},
and efficient conversion between photons at vastly different frequencies
\citep{Hill2012,Bagci2014,Andrews2015}. To realize coherent quantum
operations, the mechanical resonator needs to be prepared in the ground
state and the photon-phonon coupling rate must overcome the energy
dissipation rate of each individual system to its local environment,
reaching the so-called strong coupling regime \citep{Teufel2011,Verhagen2012,Zhang2014,Tabuchi2014}.
For this reason, realizing strong coupling in a high frequency mechanical
system is desirable since higher mechanical resonant frequency ($\omega_{m}$)
translates to lower thermal phonon occupation number ($\bar{n}\approx k_{B}T/\hbar\omega_{m}$,
where $k_{B}$ is Boltzmann's constant and $T$ is the temperature)
and hence eases refrigeration conditions for reaching the ground state
($\bar{n}\ll1$). Particularly, mechanical resonators above $\sim10\,\textrm{GHz}$
can be cooled to ground state by direct dilution refrigeration without
the necessity of active cooling techniques such as feedback cooling
\citep{Poggio2007,Lee2010} and sideband cooling \citep{Chan2011,Teufel2011a,Palomaki2013a}.

In this perspective, great efforts have been dedicated to realizing
coherent cavity electro/optomechanical coupling in high frequency
regime \citep{Xiong2013,OConnell2010b,Bochmann2013,Han2014,Han2015,Cohen2015,Balram2015}.
One remarkable breakthrough in experiment has been achieved by Cleland
group \citep{OConnell2010b}, who demonstrated the ground state cooling
of a 6-GHz piezoelectric bulk acoustic resonator (BAR) and the interrogation
of its quantum states with a superconducting phase qubit. However,
the potential of this important class of high frequency mechanical
systems can only be fully explored if its quality factor (\emph{Q}
factor) is further improved; even though the intrinsic \emph{Q} factor
of the mechanical resonator in principle can exceed $10^{5}$, the
demonstrated \emph{Q} is only 260 and the phonon lifetime is limited
to $6\,\textrm{ns}$ in Ref. \citep{OConnell2010b}. Furthermore,
the wavelength of phonon at $10$ GHz is more than four orders of
magnitude smaller than that of microwave photon. This allows not only
the reduction of the device footprint but also the study of multimode
physics beyond the single mode limitation in usual microwave resonators,
where a wealth of new phenomena emerges \citep{Krimer2014,Zhang2015,Sundaresan2015,Kostylev2016,Zhang2016}. 

In this Letter, we demonstrate a high-\emph{Q} superconducting cavity
electromechanical system operating at $10\,\textrm{GHz}$ in which
a planar superconducting microwave resonator is placed over an aluminum
nitride-on-silicon (AlN-on-Si) BAR in a non-contact configuration.
By harnessing the strong piezoelectric effect of AlN \citep{Stettenheim2010,Han2014,Han2015},
we are able to strongly couple the microwave mode with an array of
acoustic thickness modes of the BAR simultaneously. Due to the non-contact-electrode
coupling scheme, our system exhibits a high mechanical quality factor
of $7.5\times10^{5}$ and a high frequency-quality factor product
($f\cdot Q$) of $7.5\times10^{15}\,\textrm{Hz}$ at $1.7\,\textrm{K}$,
improving the phonon lifetime to $11\,\mu\textrm{s}$ at $10\,\textrm{GHz}$.
Benchmark phenomena of a strongly coupled system including avoided
crossing spectra and coherent temporal oscillations are observed,
featuring a high cooperativity of $C\approx2178$. Although our experiments
are conducted at $1.7\,\textrm{K}$, further refrigeration to millikelvins
would bring this high-performance multimode system to the quantum
regime, providing a new route for studying the complex quantum dynamics
such as entanglement between mechanical modes \citep{Schmidt2012,Seok2013}.

\emph{Cavity Piezo-electromechanical Coupling.- }The piezoelectric
interaction can be characterized by the extra electric energy due
to the strain induced polarization $H_{piezo}=\int(\Delta\mathbf{P}\cdot\mathbf{E})\mathrm{d}V=\int[(\mathbf{e}\cdot\mathbf{S})\cdot\mathbf{E}]dV$,
where $\mathbf{P}$ and $\mathbf{E}$ are the electric polarization
and the electric field vectors in the material, respectively, $\mathbf{S}$
is the second-rank strain tensor, $\mathbf{e}$ is the third-rank
piezoelectric coefficient tensor of the material. In contrast to other
electromechanical coupling mechanisms, such as capacitive coupling
and electrostriction where the interaction depends on $|\mathbf{E}|^{2}$,
the piezoelectric effect provides direct linear coupling between the
electrical field and the mechanical motion, which in the following
we refer as ``piezo-electromechanical'' coupling.

\begin{figure}
\begin{centering}
\includegraphics[width=1\columnwidth]{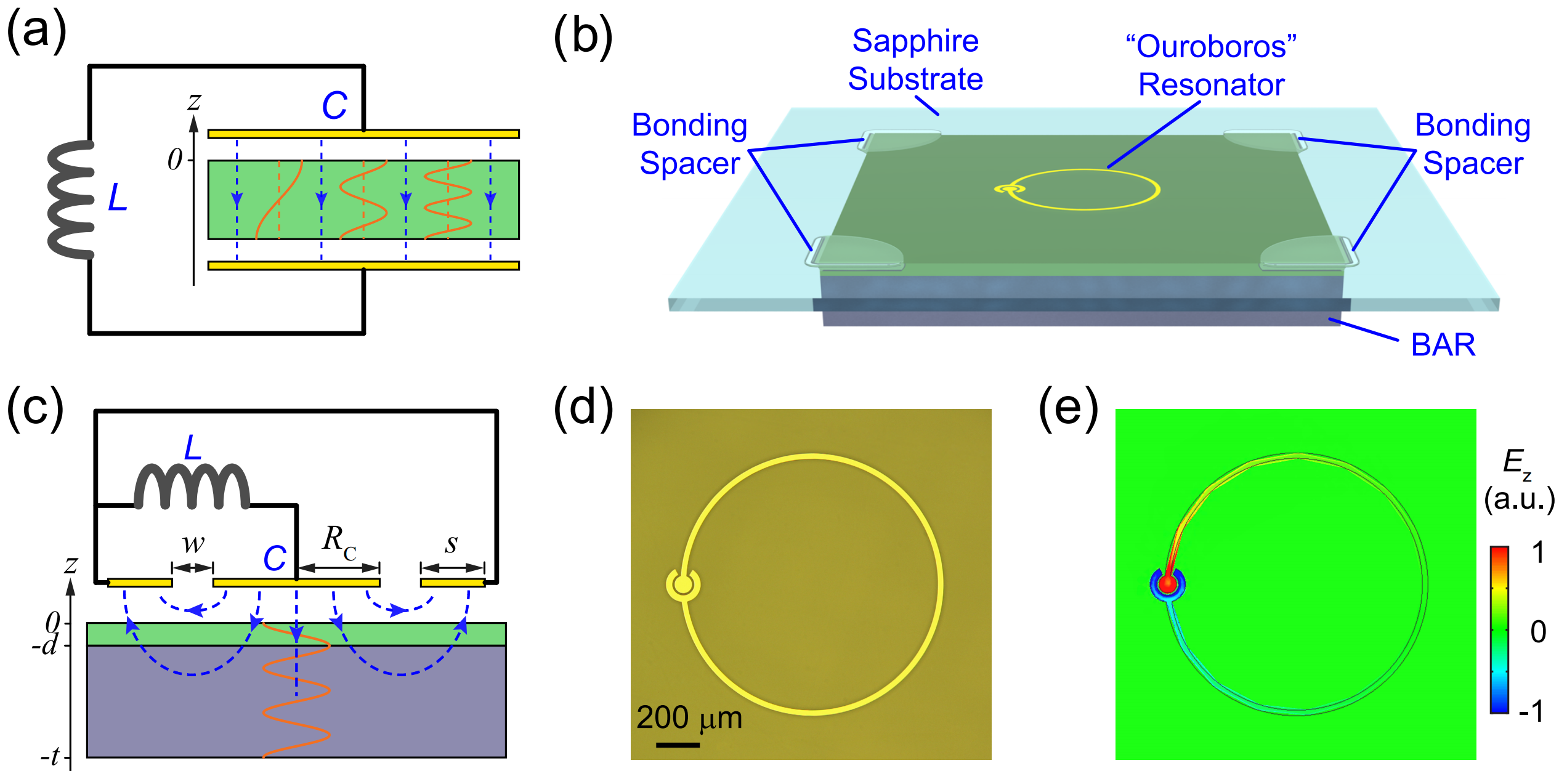}
\par\end{centering}

\caption{(a) An illustrative configuration of a cavity piezo-electromechanical
system where a piezoelectric BAR (green) is sandwiched between the
parallel capacitor plates (yellow) of an LC resonator. Blue dashed
lines: the electric field lines. Orange lines: amplitude distributions
of the \emph{z}-component mechanical displacement of different orders
of acoustic thickness modes (only the first three orders of modes
are drawn). (b) A schematic of the piezo-electromechanical device.
(Not to scale). The planar superconducting LC resonator (``Ouroboros'')
is fabricated in a 50-nm-thick niobium titanium nitride film on a
127-$\mu\textrm{m}$-thick sapphire substrate. The sapphire chip is
flipped over and suspended on top of the BAR by the bonding spacers
at the four corners of the chip. The BAR consists of a thin aluminum
nitride layer (green) deposited on top of a thick oxidized high-resistivity
silicon substrate. (c) A cross-sectional view of the device with the
``inductor'' of the ``Ouroboros'' indicated as an equivalent circuit.
The device parameters are as following: $R_{C}=40\,\textrm{\ensuremath{\mu}m}$,
$s=30\,\textrm{\ensuremath{\mu}m}$, $w=10\,\textrm{\ensuremath{\mu}m}$,
$d=550\,\textrm{nm}$, $t=500\,\textrm{\ensuremath{\mu}m}$. (d) An
optical micrograph of the \textquotedblleft Ouroboros\textquotedblright{}
resonator. The long arc-shape ``inductor'' wire has a width of $25\,\textrm{\ensuremath{\mu}m}$
and an average bending radius of $593\,\mu\textrm{m}$. (e) Simulated
mode profile of perpendicular electric field ($E_{z}$) of the \textquotedblleft Ouroboros\textquotedblright . }

\label{fig-1}
\end{figure}

To illustrate the concept of cavity piezo-electromechanical coupling,
we consider a piezoelectric film BAR sandwiched between the parallel
capacitor plates of an ``inductor-capacitor'' (LC) resonator {[}Fig.$\,$\ref{fig-1}(a){]}.
Thickness modes of the BAR form standing longitudinal acoustic waves
with the mechanical displacement sinusoidally distributed in \emph{z}-direction
{[}orange lines in Fig.$\,$\ref{fig-1}(a){]}. Due to the piezoelectric
effect, the oscillating voltage on the capacitor will compress and
expand the film, thus actuating the mechanical motion; conversely,
the strain in the film will induce excess electric polarization in
the dielectric, hence producing oscillating voltage and current in
the LC resonator. It should be noted that the coupling between the
LC resonator and the piezoelectric BAR do not require electrodes to
directly contact with the BAR. In fact, contactless coupling is preferred
because contacting metals and metal-dielectric interfaces are known
to cause mechanical dissipation \citep{Frangi2013}. The non-contact
excitation mechanism has also been recently explored in air-gap piezoelectric
MEMS devices \citep{Hung2015}. The elimination of metal from piezoelectric
structures also allows the construction of high-\emph{Q} optomechanical
resonators without metal induced light absorption, making it possible
to simultaneously couple acoustic modes to both microwave and optical
photons for realizing microwave-to-optical conversions \citep{Bochmann2013,Andrews2014}.

The linear piezo-electromechanical system can be described by a Hamiltonian
of coupled modes under the rotating-wave approximation as 
\begin{equation}
H/\hbar=\omega_{a}a^{\text{\ensuremath{\dagger}}}a+\underset{n}{\sum}\omega_{bn}b_{n}^{\text{\ensuremath{\dagger}}}b_{n}+\underset{n}{\sum}g_{n}(a^{\text{\ensuremath{\dagger}}}b_{n}+ab_{n}^{\text{\ensuremath{\dagger}}}),\label{eq:1}
\end{equation}
where $a\,(a^{\dagger})$ and $b_{n}\,(b_{n}^{\dagger})$ are the
annihilation (creation) operators for microwave photons at frequency
$\omega_{a}$ and phonons of the \emph{n}-th order mechanical mode
at frequency $\omega_{bn}$, respectively, and $g_{n}$ is the coupling
strength between them. In this work, we focus on acoustic thickness
modes with the \emph{zz}-component strain ($S_{zz}$) piezoelectrically
coupled to the \emph{z}-component electric field ($E_{z}$). Then
$g_{n}$ can be expressed as \citep{SupplementalMaterial} 
\begin{equation}
g_{n}=\frac{e_{33}}{2\sqrt{\epsilon_{0}\rho}}\sqrt{\frac{\omega_{a}}{\omega_{bn}}}\int_{V_{piezo}}{\displaystyle \zeta_{z}(\mathbf{r})}\frac{\partial}{\partial z}\xi_{nz}(\mathbf{r})\mathrm{d}V,\label{eq:2}
\end{equation}
where $e_{33}$ is the 33-component of the piezoelectric coefficient
under the contracted Voigt notation, $\epsilon_{0}$ is the vacuum
permittivity, and $\rho$ is the mass density of the mechanical resonator.
$\zeta_{z}$ and $\xi_{nz}$ are the \emph{z}-components of the normalized
electric mode profile $\boldsymbol{\zeta}(\mathbf{r})$ and the mechanical
displacement mode profile $\boldsymbol{\xi}(\mathbf{r})$ satisfying
$\int\epsilon_{r}(\mathbf{r})|\boldsymbol{\zeta}(\mathbf{r})|^{2}dV=1$
and $\int|\boldsymbol{\xi}_{n}(\mathbf{r})|^{2}dV=1$, respectively,
with $\epsilon_{r}$ being the relative dielectric constant. The integral
takes place within the volume of the piezoelectric material $V_{piezo}$.
By carefully engineering the cavity and the BAR structures to optimize
the piezoelectric mode overlap, strong electromechanical coupling
between the microwave mode and acoustic thickness modes is feasible
\citep{Zou2,SupplementalMaterial}.

\emph{Device Design.-} Since the free spectral range (FSR) of the
thickness modes is inversely proportional to the thickness of the
BAR, we utilize a thick BAR to reduce the FSR so that multiple acoustic
modes can be accessed simultaneously. The BAR consists of a thin piezoelectric
layer of \emph{c}-axis-oriented AlN deposited on top of a thick oxidized
high-resistivity Si substrate, which determines an $\textrm{FSR}\approx\frac{v_{Si}}{2t}\approx9.2\,\textrm{MHz}$,
where $v_{Si}=9.2\,\textrm{km/s}$ is the longitudinal acoustic wave
velocity in Si and $t=500\,\textrm{\ensuremath{\mu}m}$ is the thickness
of Si. The maximum coupling strength can be achieved when the thickness
of the piezoelectric layer matches half acoustic wavelength \citep{SupplementalMaterial}.
We therefore choose the thickness of the AlN layer to be $d=550\,\textrm{nm}$
to optimize the coupling strength at $\frac{\omega_{bn}}{2\pi}\approx10\,\mathrm{GHz}$
\citep{Han2015}, corresponding to mode order number $n\approx1100$.
Compared with a thin-film BAR made of pure AlN, the penalty of using
the high-order modes of the thick BAR is the reduction of the coupling
strength by a factor of $1/\sqrt{n}$. It should be noted that even
the whole BAR is made by AlN, the $1/\sqrt{n}$ scaling remains since
for high-order modes the contributions of strains with different signs
in the integral in Eq.$\,$(\ref{eq:2}) cancel each other when microwave
field doesn't vary much across the thickness of the BAR.

To optimize the electromechanical coupling strength, the electric
field of the microwave mode should be confined around the piezoelectric
layer with its lateral mode distribution matching that of the acoustic
mode. Therefore, we design a planar superconducting LC resonator so
that the AlN layer of the BAR can approach close proximity of the
capacitor surface where the electric field is concentrated. A schematic
of our device is shown in Fig.$\,$\ref{fig-1}(b). The superconducting
resonator is fabricated from a 50-nm-thick niobium titanium nitride
(NbTiN) film (critical temperature $T_{c}\approx13\,\textrm{K}$)
deposited on a 127-$\mu\textrm{m}$-thick sapphire substrate. The
sapphire chip is then flipped over and suspended on top of the BAR
using thin bonding spacers located at the four corners of the BAR.
A tiny gap of a few microns is maintained between the sapphire and
the AlN surfaces to minimize mechanical contact loss. Since the shape
of our superconducting resonator resembles the ancient Greek symbol
of a serpent eating its own tail, hereafter we name it as the \textquotedblleft Ouroboros\textquotedblright .

An optical micrograph of the \textquotedblleft Ouroboros\textquotedblright{}
resonator is shown in Fig.$\,$\ref{fig-1}(d). The structure is patterned
using e-beam lithography with hydrogen silsesquioxane (HSQ) resist
followed by chlorine-based dry etching. The \textquotedblleft Ouroboros\textquotedblright{}
consists of a ``capacitor'' formed by a central circular pad and
a surrounding ``open-ring'' pad, which are shunted by an ``inductor''
made of a long arc-shape narrow wire. The parameters of the \textquotedblleft Ouroboros\textquotedblright{}
{[}labeled in Fig.$\,$\ref{fig-1}(c){]} are designed using a finite
element high frequency simulation software (HFSS) to determine the
microwave resonance to be $\frac{\omega_{a}}{2\pi}=10\,\textrm{GHz}$
in presence of the dielectric loading of the BAR. When the ``Ouroboros''
is excited on resonance, electric field ($E_{z}$) is concentrated
and tightly confined near the capacitor pad surfaces {[}Fig.$\,$\ref{fig-1}(e){]},
improving the mode overlap in \emph{z}-direction between the microwave
mode and the AlN layer. Moreover, due to the ``$\delta v/v$'' effect
\citep{YamadaShimizu1987}, the proximal capacitor modifies the acoustic
wave velocity in the BAR and creates an effective potential well in
the lateral direction just under the location of capacitor, providing
lateral confinement for the acoustic thickness modes. Detailed discussion
on the ``$\delta v/v$'' effect induced acoustic mode bounding is
beyond the scope of this article and will be presented elsewhere \citep{Zou1}.
It is worth mentioning that another advantage of the ``Ouroboros''
design is that the long-arc inductor allows supercurrents to generate
magnetic flux far-extended from the chip surface, making it feasible
to inductively couple the \textquotedblleft Ouroboros\textquotedblright{}
with an off-chip loop probe for microwave signal input and readout.
In this way, the external microwave coupling rate of the system can
be easily tuned and optimized by changing the probe position and orientation.

\begin{figure}
\begin{centering}
\includegraphics[width=1\columnwidth]{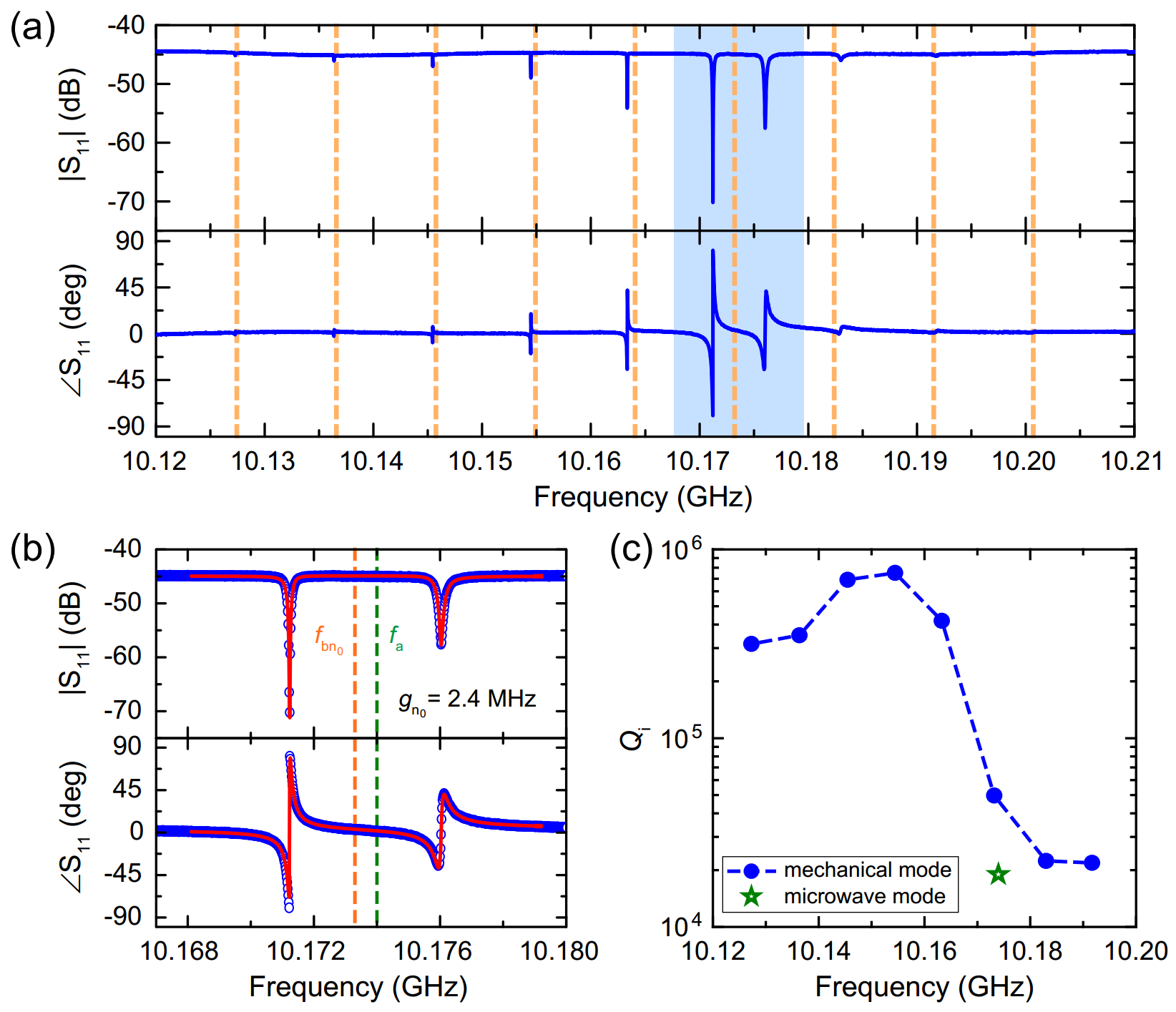}
\par\end{centering}

\caption{(a) Microwave reflection spectrum of the device measured at $1.7\,\textrm{K}$.
Orange dashed lines: unperturbed frequencies of the acoustic thickness
modes of the BAR. (b) Close-up spectrum of the regime where the microwave
mode strongly hybridizes with the acoustic thickness modes. Red lines:
fittings using the coupled-mode formula. Green and orange dashed lines:
fitted center resonant frequencies of the uncoupled microwave mode
$f_{a}=\frac{\omega_{a}}{2\pi}$ and the thickness mode $f_{bn_{0}}=\frac{\omega_{bn_{0}}}{2\pi}$,
respectively. (c) Fitted intrinsic quality factors of the microwave
mode and the different orders of acoustic thickness modes of the BAR.}

\label{fig-2}
\end{figure}
\emph{Strong Coupling.-} In order to experimentally investigate the
coherent cavity piezo-electromechanical coupling, the device is encapsulated
in a high-conductivity copper box and loaded in a close-loop refrigerator
with a base temperature of $1.62\,\textrm{K}$ \citep{Wang2014}.
In all experiments, the power of the microwave probe signal is set
to be below $-60\,\mathrm{dBm}$ to avoid undesired nonlinear and
heating effects in the superconducting resonator.

We first study the properties of the coupled system by probing the
microwave reflection spectrum using a vector network analyzer. Figure$\,$\ref{fig-2}(a)
shows the amplitude and the phase spectra of the reflection coefficient
$S_{11}$ at $1.7\,\textrm{K}$. Several sharp resonance dips with
distinct phase changes are clearly observed. Among all the resonances,
the two most prominent dips at $10.171\,\textrm{GHz}$ and $10.176\,\textrm{GHz}$
are particularly close to each other {[}light blue area in Fig.$\,$\ref{fig-2}(a){]}.
As the resonances become further away from these two center resonances,
the mode spacing becomes more uniform and gradually approaches a constant
of $9.1\,\textrm{MHz}$, which matches well with the calculated FSR
of the acoustic thickness modes of the BAR. This implies that those
\textquotedblleft side bands\textquotedblright{} originate from thickness
modes of different orders, whereas the \textquotedblleft Ouroboros\textquotedblright{}
resonance is mostly hybridized with a closely matched thickness mode
($n=n_{0}$) to produce the two hybrid resonances at $10.171\,\textrm{GHz}$
and $10.176\,\textrm{GHz}$. To better visualize the coupling induced
frequency shifts, we measure the frequency difference between two
far-separated resonances away from $10.173\,\textrm{GHz}$, and count
the number of mechanical resonances between them. The ratio gives
the FSR of the thickness modes which is found to be $9.12\,\textrm{MHz}$.
Based on this measured FSR value, we overlay the unperturbed frequencies
of thickness modes as vertical orange dashed lines in Fig.$\,$\ref{fig-2}(a).
It be can seen that the density of modes is mostly modified around
the two resonances at $10.171\,\textrm{GHz}$ and $10.176\,\textrm{GHz}$
where the electromechanical coupling leads to strong hybridization
of the microwave and the mechanical modes.

The spectra of the hybridized modes are fitted {[}Fig.$\,$\ref{fig-2}(b){]}
by
\begin{equation}
S_{11}(\omega)=-1+\frac{2\kappa_{a,e}}{-i(\omega-\omega_{a})+\kappa_{a,i}+\kappa_{a,e}-\underset{n}{\Sigma}\frac{|g_{n}|^{2}}{i(\omega-\omega_{bn})-\kappa_{bn}}},\label{eq:3}
\end{equation}
which is the steady-state solution of the coupled-multimode formula.
Here, $\kappa_{a,i}$ and $\kappa_{a,e}$ are the intrinsic dissipation
rate and the external coupling rate of the ``Ouroboros'', respectively.
As expected, the fitted frequencies of the uncoupled microwave mode
$\frac{\omega_{a}}{2\pi}=10.1740\,\mathrm{GHz}$ and the thickness
mode $\frac{\omega_{bn_{0}}}{2\pi}=10.1733\,\mathrm{GHz}$ are very
close to each other. The coupling strength is extracted to be $\frac{g_{n_{0}}}{2\pi}=2.4\,\mathrm{MHz}$,
larger than the dissipation rates of both the microwave mode $\frac{\kappa_{a,i}+\kappa_{a,e}}{2\pi}=0.39\,\mathrm{MHz}$
and the thickness mode $\frac{\kappa_{bn_{0}}}{2\pi}=0.10\,\mathrm{MHz}$.
Therefore, our system is in the strong coupling regime \citep{Tabuchi2014,Zhang2014}
and a cooperativity of $C=g_{n_{0}}^{2}/(\kappa_{a}\kappa_{bn_{0}})\approx148$
is obtained.

By this method, the intrinsic losses of all the thickness modes in
the spectrum in Fig.$\,$\ref{fig-2}(a) are extracted and the corresponding
\emph{Q}-factors are plotted in Fig.$\,$\ref{fig-2}(c). A highest
mechanical \emph{Q}-factor of $7.5\times10^{5}$ is obtained, giving
rise to an ultra-high cooperativity of $C\approx2178$. An important
figure of merit quantifying the decoupling of a mechanical resonator
from the thermal environment is the $f\cdot Q$ product \citep{Aspelmeyer2014,Chang2012}.
Our system exhibits a high $f\cdot Q$ product of $7.5\times10^{15}\,\textrm{Hz}$
at $1.7\,\textrm{K}$, more than three orders larger than previous
experimental result \citep{OConnell2010b}. This high $f\cdot Q$
product translates to excellent coherence characterized by  a large
number of coherent oscillations given by $\frac{hfQ}{k_{B}T}\approx2\times10^{5}$
in presence of thermal decoherence. The asymmetry of the mechanical
\emph{Q}-factors with respect to the ``Ouroboros'' resonant frequency
can be attributed to the microwave cavity-mediated coupling between
the localized thickness modes in the potential well and the unlocalized
leaky modes \citep{Zou1}.

\begin{figure}
\begin{centering}
\includegraphics[width=1\columnwidth]{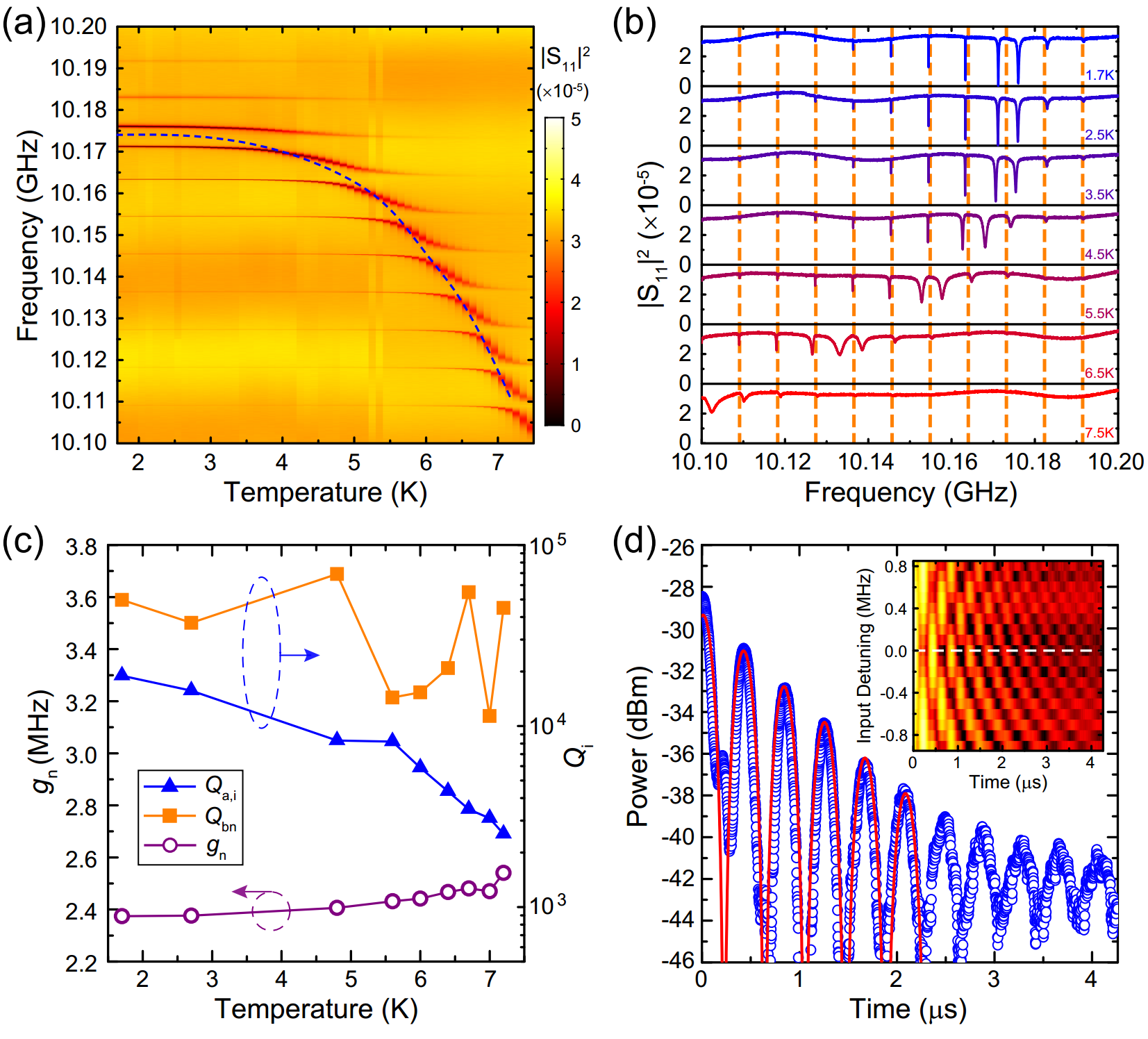}
\par\end{centering}

\caption{(a) Temperature dependence of the microwave reflection spectra. Blue
dashed line: the uncoupled ``Ouroboros'' resonant frequency obtained
from the fittings using the coupled-mode formula. (b) Line plots of
the microwave reflection spectra at different temperatures. Orange
dashed lines: unperturbed frequencies of the acoustic thickness modes
of the BAR. (c) Coupling strength and intrinsic \emph{Q}-factors extracted
at temperatures when the ``Ouroboros'' resonance is tuned across
a thickness mode. (d) Microwave reflection in temporal domain. The
center frequency of the input microwave pulse is $10.1737\,\textrm{GHz}$.
Red line: the fitting of the exponentially decaying oscillation. Inset:
temporal oscillations at different input frequencies detuned from
$10.1737\,\textrm{GHz}$ (white dashed line).}

\label{fig-3}
\end{figure}

\emph{Temperature Dependence and Frequency Tuning.-} The features
of multimode strong coupling in our system are further investigated
by gradually sweeping the ``Ouroboros'' resonant frequency via temperature
tuning. When the temperature increases, the \textquotedblleft Ouroboros\textquotedblright{}
resonant frequency decreases monotonically due to the reduction of
the density of cooper pairs in the superconductor. Figure$\,$\ref{fig-3}(a)
shows the microwave reflection spectra with the temperature varying
from 1.7 K to 7.5 K. As the ``Ouroboros'' resonant frequency is
swept over several FSRs of the thickness modes, avoided crossings
are observed in sequence as a signature of multimode strong coupling
(MMSC) \citep{Sundaresan2015,Noguchi2016}. In Fig.$\,$\ref{fig-3}(b),
typical spectra at different temperatures are shown in linear plot.
It can be seen that at higher temperatures the coupled modes shift
dramatically to lower frequencies with increased resonance linewidths
and reduced dip extinctions due to increased internal dissipation.

Using the coupled-multimode model, the uncoupled \textquotedblleft Ouroboros\textquotedblright{}
resonant frequency at different temperatures is extracted and plotted
as the blue dashed line in Fig.$\,$\ref{fig-3}(a), which agrees
with the kinetic inductance model of thin-film superconducting resonators
\citep{Gao2008}. The BAR mechanical resonant frequencies, on the
other hand, shift less than $5\,\textrm{Hz}$ because of the extremely
small thermal expansion coefficient of silicon within the temperature
range of our experiments \citep{Swenson1983}. Figure$\,$\ref{fig-3}(c)
shows the coupling strength and the intrinsic \emph{Q}-factors extracted
at temperatures when the ``Ouroboros'' resonance is swept across
a thickness mode. In the temperature range we studied, $g_{n}$ remains
almost constant around $2.4\,\textrm{MHz}$. The intrinsic \emph{Q}-factor
of the ``Ouroboros'' gradually drops due to the increasing quasi-particle
dissipation, while the \emph{Q}-factors of the thickness modes show
a general decreasing trend but with fluctuations, which could be attributed
to property variations among different orders of modes and subject
of further investigations.

\emph{Dynamics.-} The strong coupling permits coherent energy exchange
between coupled modes. We therefore study the dynamic interaction
between the microwave and the mechanical modes in time domain. Experimentally,
a short microwave excitation pulse is injected into the coupled system
and the reflected power is monitored at $1.7\,\textrm{K}$. Periodical
oscillations in time domain are observed with their periods depending
on the center frequency of the input pulse {[}inset of Fig.$\,$\ref{fig-3}(d){]}.
The most distinct oscillation with smallest period is observed at
$10.1737\,\textrm{GHz}$ {[}white dashed line in Fig.$\,$\ref{fig-3}(d)
inset{]}, which matches well with the average frequency between the
``Ouroboros'' and the closely coupled mechanical resonance. The
corresponding time trace is plotted in Fig.$\,$\ref{fig-3}(d). Because
of the fast coherent energy exchange between the strongly coupled
microwave and mechanical modes, the reflected power oscillates periodically
with exponentially decaying amplitude. The oscillation period is fitted
to be $4.1\,\textrm{\ensuremath{\mu}s}$ {[}red line in Fig.$\,$\ref{fig-3}(d){]},
showing excellent agreement with the measured coupling strength $2\pi/g_{n}=4.2\,\textrm{\ensuremath{\mu}s}$.
The slight deviation from the exponential decay trend for oscillation
signal after around $2\,\textrm{\ensuremath{\mu}s}$ could be attributed
to the coupling between the microwave mode and multiple neighboring
thickness modes.

\emph{Conclusion.-} We have achieved multimode electromechanical strong
coupling between a superconducting microwave resonator and a piezoelectric
BAR at $10\,\textrm{GHz}$ with a cooperativity of exceeding $2000$.
Coherent interaction has been demonstrated via the observation of
the avoided-crossing spectra and the temporal oscillation of microwave
energy. Due to the high operating frequency, our system can be expected
to incorporate with quantum superconducting circuits for studying
mechanical quantum states at relaxed refrigeration temperatures. The
achieved ultra-high $f\cdot Q$ product of $7.5\times10^{15}\,\textrm{Hz}$
and large coupling strength-over-FSR ratio of $\frac{g_{n}}{\textrm{FRS}}\approx26\%$
allow realization of mechanical multimode quantum memory devices \citep{Zhang2015}
and exploration of the intriguing superstrong coupling regime \citep{Sundaresan2015,Zhang2016}.

\textbf{Acknowledgment} C.L.Z. thanks Liang Jiang for helpful discussions.
This work is supported by Laboratory of Physical Sciences through
a grant from Army Research Office (W911NF-14-1-0563)), an Air Force
Office of Scientific Research (AFOSR) MURI grant (FA9550-15-1-0029)
and a NSF MRSEC grant (1119826). H.X.T. acknowledges support from
a Packard Fellowship in Sceince and Engineering. The authors thank
Michael Power and Dr. Michael Rooks for assistance in device fabrication.

\end{document}